# Possible unconventional superconductivity in iron-based layered compound LaFePO: Study of heat capacity

Yoshimitsu Kohama[1], Yoichi Kamihara[2], Hitoshi Kawaji[1], Tooru Atake[1], Masahiro Hirano[2,3] and Hideo Hosono[1–3]

[1] *Materials and Structures Laboratory, Tokyo Institute of Technology*

*Mail-Box: R3-7, 4259 Nagatsuta-cho, Midori-ku, Yokohama 226-8503, Japan*

[2] *ERATO-SORST, Japan Science and Technology Agency in Frontier Research Center, Tokyo Institute of Technology*

*Mail-Box: S2-13, 4259 Nagatsuta-cho, Midori-ku, Yokohama 226-8503, Japan*

[3] *Frontier Research Center, Tokyo Institute of Technology*

*Mail-Box: S2-13, 4259 Nagatsuta-cho, Midori-ku, Yokohama 226-8503, Japan*

Heat capacity measurements were performed on recently discovered, iron-based layered superconductors, non-doped LaFePO and fluorine-doped LaFePO ($LaFePO_{0.94}F_{0.06}$). A relatively large electronic heat capacity coefficient ($\gamma$) of 10.6 mJ $K^{-2}$ $mol^{-1}$ and a small normalized heat capacity jump ($\Delta C_p/\gamma T_c$) of 0.29 at $T_c$ = 3.3 K were observed in LaFePO. $LaFePO_{0.94}F_{0.06}$ had a smaller $\gamma$ of 8.3 mJ $K^{-2}$ $mol^{-1}$ and a larger $\Delta C_p/\gamma T_c$ of 0.44 at $T_c$ = 5.8 K. These values indicate that these compounds have strong electron–electron correlation and magnetic spin fluctuation, which are the signatures of unconventional superconductivity mediated by spin fluctuation.

## 1. Introduction

Following the discovery of superconductivity in a high-$T_c$ cuprate by Bednorz and Müller,[1] several layered superconductors have been found in ruthenates,[2] borocarbides,[3] organic compounds,[4] sodium cobalt oxyhydrate,[5] etc. The most peculiar feature commonly seen in these layered compounds is that the superconductivity is mediated by an unconventional pairing mechanism. Recently, superconductivity has been discovered in a family of layered oxypnictide, LaFePO ($T_c \sim$ 7 K)[6] and LaFeAsO ($T_c \sim$ 26 K)[7]. The crystal structure of LaFe$Pn$O ($Pn$ = P, As) is tetragonal (space group $P4/nmm$, $a = b = 0.3964(1)$ and $c = 0.8512(3)$ nm for LaFePO)[6] and is composed of alternate stacking of iron pnictide (Fe$Pn$) and lanthanum oxide (LaO) layers. The Fe$Pn$ layers, which consist of a two-dimensional array of edge-sharing Fe$Pn_4$ tetrahedrons, are regarded as conduction layers.[8-10] On the other hand, the LaO layers are insulating layers and play the role of an electron donor to Fe$Pn$ layers. The alternate stacking of superconducting Fe$Pn$ and insulating LaO layers is reminiscent of the stack structure of superconducting CuO$_2$ planes and insulating blocks in high $T_c$-cuprates. In addition, recent photoemission spectroscopy and band calculation[8] have clarified that the Fe $3d$ and P $3p$ states mainly contribute to the electronic states around the Fermi level, which is similar to the hybridized state of the CuO$_2$ plane made by the Cu $3d$ and O $2p$ states. These structural and electronical similarities between LaFePO and high-$T_c$ cuprates may help explain the unconventional characteristics of the transition metal-based superconductors. However, it is still an open question whether LaFePO is an unconventional superconductor, and its clarification is an important subject for the study of LaFePO systems.

In analogy with high-$T_c$ cuprates, the systematic study of doping—Ca substitution for La (hole doping)[8] and F substitution for O (electron doping)[5,8]—was performed in an LaFePO system, and the highest $T_c$ sample ($T_c \sim$ 7 K) was obtained in F-doped LaFePO.[8] In this study, we evaluated the heat capacity of non-doped and F-doped LaFePO at temperatures from 0.085 to 300 K. The results strongly suggest an unconventional pairing mechanism.

## 2. Experimental

Polycrystalline samples of LaFePO (non-doped LaFePO) and LaFePO$_{0.94}$F$_{0.06}$ (F-doped LaFePO) were synthesized by conventional solid-state reactions, as described in a previous paper.[6] X-ray analysis indicated that both compounds had a tetragonal structure, in which the *a*-axis of F-doped LaFePO was smaller than that of non-doped LaFePO by only ~0.15%, as described in our previous report.[8]

Heat capacity measurements were performed using two different calorimeters employing a relaxation method. A homemade calorimeter with a dilution refrigerator was used for the heat capacity measurements in the temperature region from 0.085 to 5 K.[11] Quantum Design PPMS was used for the measurements in the temperature region from 2 to 300 K.

## 3. Results and discussion

The heat capacity ($C_p$) of non-doped and F-doped LaFePO below 8 K is shown in Fig. 1, and the inset shows variation in $C_p$ over the whole temperature region. The $C_p$ curves of both compounds above 8 K are in good agreement with each other, indicating a similar lattice heat capacity ($C_{lat}$). In contrast, the $C_p$ curves in the low-temperature region are significantly different. The difference should relate to the electronic heat capacity ($C_{ele}$), which has a dominant contribution at low temperatures.

In this low-temperature region, $C_{lat}$ follows the Debye law (i.e., $C_{lat} \sim T^3$) in both the normal and superconducting states, and $C_{ele}$ varies linearly with $T$ in the normal state. Thus, the total heat capacity in the normal state can be expressed as $C_p = C_{lat} + C_{ele} = \beta T^3 + \gamma T$, where $\beta$ and $\gamma$ are the lattice and electronic heat capacity coefficients, respectively. Figure 2 shows the plots of $C_p T^{-1}$ versus $T^2$ below 8 K. Here, the linear dependence of $C_p T^{-1}$ on $T^2$, indicating the normal state, is observed in non-doped LaFePO above ~5 K and in F-doped

LaFePO above ~7 K. On the other hand, a hump associated with superconducting transition is observed at $T_c$ = 3.3 K for non-doped LaFePO and at $T_c$ = 5.8 K for F-doped LaFePO. The peak temperatures are consistent with the $T_c$ values reported previously.[6,8] Under magnetic field, the hump of F-doped LaFePO is not observed, and a Schottky-like anomaly appears below 2 K, as shown in the inset of Fig. 2. The peak temperature of the Schottky-like anomaly increases with increasing magnetic field. Such magnetic field dependence of a Schottky-like anomaly is usually observed in a system containing paramagnetic impurity,[11] because of the Zeeman effect of impurity spin. Assuming that the degeneracy of the impurity-spin energy levels ($D$) is 2, the concentration of paramagnetic impurities ($n$) is estimated to be $4.4 \times 10^{-4}$ mol$^{-1}$ mol from the height of the Schottky anomaly at 1 T. If $D$ is larger than 2, the estimated value of concentration is further reduced (e. g., $n = 3.0 \times 10^{-4}$ for $D = 3$, $n = 2.6 \times 10^{-4}$ for $D = 4$, $n = 2.4 \times 10^{-4}$ for $D = 5$, $n = 2.3 \times 10^{-4}$ mol$^{-1}$ mol for $D = 6$). Since the La metal source often includes a small amount of Ce metal, the paramagnetic impurity may be $Ce^{3+}$ ions ($J = 5/2$). $D$ of the $Ce^{3+}$ ground state can be changed from 2 to 6 depending on the crystal field, resulting in an $n$ value of $2.3 \times 10^{-4}$ to $4.4 \times 10^{-4}$ mol$^{-1}$ mol. The magnetic impurities act as scattering centers and promote dramatic effects called pair-breaking effects, which lead to suppression of the $T_c$ and the heat capacity jump at $T_c$.[12-16] Although the observed amount of impurity is considerably small compared with the impurity concentration ($n = 1 \times 10^{-2}$ mol$^{-1}$ mol) that induces complete loss of superconductivity in phonon-mediated BCS superconductors,[17] a little impurity may significantly suppress the superconducting transition because of the strong pair-breaking effect, as seen in some unconventional superconductors such as $Sr_2RuO_4$[14–16,18] and $UPt_3$.[19]

    The intercept of the straight lines in Fig. 2 corresponds to the electronic heat capacity coefficient $\gamma$, and the slope corresponds to the lattice heat capacity coefficient $\beta$. Since the Schottky-like anomaly significantly decreases with increasing temperature, the heat capacity under a magnetic field of 2 T above 3 K is regarded as the heat capacity of the normal state.

As a result, the parameters in both compounds (estimated by a standard least-squares method) are $\beta = 0.15$ mJ K$^{-4}$ mol$^{-1}$ and $\gamma = 10.6$ mJ K$^{-2}$ mol$^{-1}$ for non-doped LaFePO, and $\beta = 0.17$ mJ K$^{-4}$ mol$^{-1}$ and $\gamma = 8.3$ mJ K$^{-2}$ mol$^{-1}$ for F-doped LaFePO. The Debye temperature ($\Theta_D$) calculated from $\Theta_D^3 = (12/5)\pi^4 R(r/\beta)$, where $r$ is the number of atoms in the chemical formula and $R$ is the gas constant, is 371 and 361 K for non-doped and F-doped LaFePO, respectively. The closeness in the $\Theta_D$ values of the two compounds indicates that F-doping barely affects lattice vibration. On the other hand, the magnitude of $\gamma$ in both compounds reflects the electronic state and is comparable with that of $d$-band metals (Fe: 5.02 mJ K$^{-2}$ mol$^{-1}$, Ni: 7.28 mJ K$^{-2}$ mol$^{-1}$, Co: 4.75 mJ K$^{-2}$ mol$^{-1}$).[20] This suggests a major contribution of $d$-band metals to the electronic density of states at the Fermi surface ($N_D^*$) in an LaFePO system, as indicated by the previous band calculation.[8,10] The electronic density of states is calculated to be $6.73 \times 10^{22}$ eV$^{-1}$ cm$^{-3}$ (4.50 states eV$^{-1}$) and $5.26 \times 10^{22}$ eV$^{-1}$ cm$^{-3}$ (3.50 states eV$^{-1}$) for non-doped and F-doped LaFePO, respectively, using $\gamma = (1/3)\pi^2 k_B^2 N_D^*$, where $k_B$ is the Boltzmann constant. It is noteworthy that the value of $N_D^*$ is clearly larger than the band density of states ($N_D$) obtained by the previous band calculation ($N_D = 1.71$ states eV$^{-1}$ for non-doped LaFePO).[8] This enhancement can be described in terms of the ratio of the effective mass ($m_{eff}$) to the band mass ($m_B$), which is given by $m_{eff}/m_B = N_D^*/N_D$. The estimated $m_{eff}/m_B$ of 2.63 in non-doped LaFePO is comparable to that of the strongly correlated superconductor Sr$_2$RuO$_4$ ($m_{eff}/m_B = 3\sim 4$),[21,22] and clearly indicates the strong electron–electron correlation in LaFePO systems. If the mass enhancement results from electron–phonon interactions causing BCS superconductivity, the extraordinarily high $T_c$ of ~35 K can be calculated from the well-known McMillan equation,[23] assuming a Coulomb pseudo-potential of 0.12. Considering the low superconducting transition temperature, it is not plausible to speculate that electron–phonon interactions provide the major contribution to the mass enhancement. On the other hand, it is important that the $N_D^*$ of the F-doped LaFePO that has a higher $T_c$ (5.8 K) is smaller than that of the non-doped LaFePO. The ratio of $N_D^*$ in F-doped LaFePO to that in non-doped LaFePO is ~0.8, which is compatible with the reported ratio of 0.7 estimated by NMR measurements.[9] According to the BCS theory, the transition temperature should increase with increase in the density of states.[23] Thus, the opposite relationship between the density of states and $T_c$ suggests the unconventional mechanism of superconductivity in the present system.

$C_{ele}$ obtained by subtracting the lattice contribution ($\beta T^3$) from the total $C_p$ is shown in Fig. 3 using the plots of $C_{ele}/T$ versus $T$. Assuming a discontinuous entropy-conserved superconducting transition, we observe heat capacity jumps at $T_c$ ($\Delta C_p/T_c$) of 3.1 and 3.6 mJ mol$^{-1}$ K$^{-2}$ for non-doped and F-doped LaFePO, respectively. The normalized heat capacity jump, $\Delta C_p/\gamma T_c$, is 0.29 and 0.44 for non-doped and F-doped LaFePO, respectively. The smaller value of $\Delta C_p/\gamma T_c$ in non-doped LaFePO may indicate a stronger pair-breaking effect because of the higher concentration of impurity.

Below 0.4 K, $C_{ele}/T$ in F-doped LaFePO shows a tendency to saturate to a finite intercept at 0 K. The intercept corresponds to a residual $T$-linear coefficient ($\gamma_0$) of 4.1 mJ K$^{-2}$ mol$^{-1}$, which is 50% of $\gamma$. $\gamma_0$ is usually observed in a system with pair-breaking effects, such as zinc-substituted high-$T_c$ cuprates[12,13] and some impurities including Sr$_2$RuO$_4$.[14-16] Since the F-doped LaFePO sample includes some impurities, the $\gamma_0$ is expected to originate from the pair-breaking effects. In the previous resistivity measurements,[8] the higher $T_c$ of 7.16 K was observed in the F-doped LaFePO sample. Assuming that the $T_c$ obtained by resistivity measurements indicates an ideal superconducting transition temperature ($T_{c0}$) without pair-breaking effects, the ratio of $T_c$ to $T_{c0}$ is $T_c/T_{c0} = 0.80$. The present results of $\Delta C_p/\gamma T_c$ (0.44), $\gamma/\gamma_0$ (0.50), and $T_c/T_{c0}$ (0.80) differ from the expected values in conventional BCS $s$-wave superconductors, in which $\Delta C_p/\gamma T_c$ is 0.75–1.11 and $\gamma/\gamma_0$ is 0–0.38 at $T_c/T_{c0} = 0.80$.[24] In $p$-wave superconductors (e.g., Sr$_2$RuO$_4$), $\Delta C_p/\gamma T_c$ is about 0.4 and $\gamma/\gamma_0$ is about 0.5 around $T_c/T_{c0}$ of 0.80.[14-16] In the case of $d$-wave superconductors such as high-$T_c$ cuprates, $\Delta C_p/\gamma T_c$ is about 0.6 and $\gamma/\gamma_0$ is about 0.5.[13] In this context, it is natural to consider that the mechanism of superconductivity associated with electrons in the FeP layers of LaFePO is quite different from that of BCS superconductors.

The temperature-independent dc magnetic susceptibility ($\chi_0$) was reported as $5.91 \times 10^{-4}$ emu/mole and $2.10 \times 10^{-4}$ emu/mole for non-doped and F-doped LaFePO, respectively.[8] The temperature-independent term is decomposed into four terms:

$\chi_0 = \chi_P + \chi_C + \chi_L + \chi_{VV}$,

where $\chi_P$, $\chi_C$, $\chi_L$, and $\chi_{VV}$ are the Pauli paramagnetism, core-electron diamagnetism, Landau diamagnetism, and van Vleck paramagnetism, respectively. The value of $\chi_L$ should be negligibly small compared with that of $\chi_P$, because of the large value of $m_{eff}/m_B$ in the present system. The $\chi_{VV}$ is also expected to be negligibly small because of the obviously small value of $\chi_{VV}$ in similar compounds.[25-27] Therefore, we obtain $\chi_P$ values of $6.7 \times 10^{-4}$ and $2.9 \times 10^{-4}$ emu/mole for non-doped and F-doped LaFePO, respectively, by correcting $\chi_C$.[28] Wilson ratios ($R_W$) of 4.6 for non-doped LaFePO and of 2.6 for F-doped LaFePO are obtained from the following relation: $R_W \equiv \left(\pi^2 k_B^2 / 3\mu_B^2\right)\left(\chi_P / \gamma\right)$, where $\mu_B$ represents the Bohr magneton. The obtained values of $R_W$ are larger than 2, which should be observed in a system with only electron–electron correlation. The present large values suggest that a LaFePO system is a ferromagnetic spin fluctuation system, similar to normal-fluid $^3$He ($R_W \sim 3.3$)[29] and $Sr_3Ru_2O_7$ ($R_W \geq 10$)[30]. The ferromagnetic spin fluctuation disfavors formation of the spin singlet by either s-wave (BCS) or d-wave pairing,[31,32] and favors formation of the spin triplet by p- or f-wave pairing, as observed in superfluid $^3$He[29] and $Sr_2RuO_4$[33]. This suggests that the LaFePO system behaves as a spin triplet superconductor, possibly because of the spin fluctuation.

Although the above discussions imply that LaFePO systems are a new class of unconventional superconductors, they do not provide direct evidence. Further investigation on single crystals is therefore needed.

**4. Summary**

The electronic heat capacity coefficient ($\gamma$), superconducting phase transition temperature ($T_c$), and the height of heat capacity jump at $T_c$ ($\Delta C_p$) were evaluated. From these data, a large mass enhancement of $m_{eff}/m_B = 2.64$ in non-doped LaFePO and a large Wilson ratio of $R_W = 2.6 \sim 4.6$ were estimated. These values imply that the electronic state in LaFePO systems can be a novel Fermi liquid state with a strong electron–electron correlation and a possible large ferromagnetic spin fluctuation. Hence, LaFePO systems seem to be an example of unconventional superconductors induced by ferromagnetic spin fluctuation.

**Acknowledgments**

The authors wish to thank Prof. Kenji Ishida of Kyoto University for stimulating discussions concerning this work. This work was partly supported by Grant-in-Aid for JSPS Fellows (No.19·9728 to Y. Kohama).

**References**

1) J. G. Bednorz, and K. A. Müller, Z. Phys. B **64**, 189 (1986).

2) Y. Maeno, H. Hashimoto, K. Yoshida, S. Nishizaki, T. Fujita, J. G. Bednorz, and F. Lichtenberg, Nature (London) **372**, 532 (1994).

3) R. J. Cava, H. Takagi, B. Batlogg, H. W. Zandbergen, J. J. Krajewski, W. F. Peck, Jr., R. B. van Dover, R. J. Felder, T. Siegrist, K. Mizuhashi, J. O. Lee, H. Eisaki, S. A. Carter, and S. Uchida, Nature (London) **367**, 146 (1994).

4) S. Belin, K. Behnia, and A. Deluzet, Phys. Rev. Lett. **81**, 4728 (1998).

5) K. Takada, H. Sakurai, E. Takayama-Muromachi, F. Izumi, R. A. Dilanian, and T. Sasaki, Nature (London) **422**, 53 (2003).

6) Y. Kamihara, H. Hiramatsu, M. Hirano, R. Kawamura, H. Yanagi, T. Kamiya, and H. Hosono, J. Am. Chem. Soc. **128**, 10012 (2006).

7) Y. Kamihara, T. Watanabe, M. Hirano, and H. Hosono, J. Am. Chem. Soc., **130** 3296 (2008).

8) Y. Kamihara, M. Hirano, H. Yanagi, T. Kamiya, Y. Kobayashi, Y. Saitoh, E. Ikenaga, K. Kobayashi, and H. Hosono, Phys. Rev. B **77**, 214515 (2008).

9) Y. Nakai et al, (unpublished).

10) S. Lebègue, Phys. Rev. B **75**, 35110 (2007).

11) Y. Kohama, T. Tojo, H. Kawaji, T. Atake, S. Matsuishi, and H. Hosono, Chem. Phys. Lett. **421**, 558 (2006).

12) N. E. Phillips, R. A. Fisher, J. E. Gordon, S. Kim, A. M. Stacy, M. K. Crawford, and E. M. McCarron, III, Phys. Rev. Lett. **65**, 357 (1990).

13) K. Maki, and H. Won, Ann. Phisik **5**, 320 (1996).

14) S. Nishizaki, Y. Maeno, and Z. Mao, J. Low Temp. Phys. **117**, 1581 (1999).

15) S. Nishizaki, Y. Maeno, S. Farner, S. Ikeda, and T. Fujita, J. Phys. Soc. Jpn. **67**, 560 (1998).

16) K. Miyake, and O. Narikiyo, Phys. Rev. Lett. **83**, 1423 (1999).

17) B. T. Matthias, H. Suhl, and E. Corenzwit, Phys. Rev. Lett. **1**, 92 (1958).

18) A. P. Mackenzie, R. K. W. Haselwimmer, A. W. Tyler, G. G. Lonzarich, Y. Mori, S. Nishizaki, and Y. Maeno, Phys. Rev. Lett. **80**, 161 (1998).

19) R. Joynt, and L. Taillefer, Rev. Mod. Phys. **74**, 235 (2002).

20) E. S. R. Gopal, Specific Heats at Low Temperatures (Heywood Books, London, 1966).

21) A. P. Mackenzie, S. R. Julian, A. J. Diver, G. J. McMullan, M. P. Ray, G. G. Lonzarich, Y. Maeno, S. Nishizaki, and T. Fujita, Phys. Rev. Lett. **76**, 3786 (1996).

22) Y. Maeno, S. Nishizaki, K. Yoshida, S. Ikeda, and T. Fujita, J. Low Temp. Phys. **105**, 1577 (1996).

23) W. L. McMillan, Phys. Rev. **167**, 331 (1968).

24) S. Takayanagi, and T. Sugawara, J. Phys. Soc. Jpn. **38**, 718 (1975).

25) For instance, the value of $\chi_0$ in insulating LaOZnP was explained by only core magnetism, and no remarkable van Vleck term was reported.[26] In addition, the value of $\chi_0$ in a similar sulfide compound, LaOCuS, is about $-0.6 \times 10^{-4}$ emu/mol,[27] and the core diamagnetism term (about $-0.8 \times 10^{-4}$ emu/mol) corrected value (about $0.2 \times 10^{-4}$ emu/mol), which may correspond to a van Vleck term, is also negligibly small.

26) Y. Takano, S. Komatsuzaki, H. Komasaki, T. Watanabe, Y. Takahashi, and K. Takase, J. Alloy. Comp., to be published.

27) K. Takase, T. Shimizu, K. Makihara, Y. Takahashi, Y. Takano, K. Sekizawa, Y. Kuroiwa, S. Aoyagi, and A. Utsumi, Physica B **329–333**, 961 (2003).

28) For extracting the intrinsic Pauli magnetic terms from previous data,[8] core diamagnetic susceptibilities were calculated using the values in literature: $-0.20 \times 10^{-4}$ for $La^{3+}$, $-0.12 \times 10^{-4}$ for O, $-0.11 \times 10^{-4}$ for $F^-$, $-0.22 \times 10^{-4}$ for $Fe^-$, $-0.26 \times 10^{-4}$ emu mol$^{-1}$ for P giving $\chi_{core}$ values of $-0.80 \times 10^{-4}$ emu mol$^{-1}$ for both non-doped and F-doped LaFePO.

29) C. Enss, and S. Hunklinger, Low-temperature Physics (Springer, Berlin, 2005).

30) S. Ikeda, Y. Maeno, S. Nakatsuji, M. Kosaka, and Y. Uwatoko, Phys. Rev. B **62**, R6089 (2000).

31) T. M. Rice, and M. Sigrist, J. Phys. Condens. Matter. **7**, L643 (1995).

32) K. Machida, M. Ozaki, and T. Ohmi, J. Phys. Soc. Jpn. **65**, 3720 (1996).

33) I. I. Mazin, and D. J. Singh, Phys. Rev. Lett. **79**, 733 (1997).

**Figure captions**

**Fig. 1.** Heat capacity ($C_p$) of non-doped and F-doped LaFePO. The solid triangles correspond to PPMS measurement of non-doped LaFePO. The solid and open circles correspond to measurement of F-doped LaFePO by PPMS and using a homemade calorimeter, respectively. The inset shows the $C_p$–$T$ curve over the whole temperature region.

**Fig. 2.** Heat capacity of non-doped and F-doped LaFePO, plotted as $C_p T^{-1}$ vs. $T^2$ for $0 < T \leq 8$ K. The dotted and dashed lines represent the linear fit of the normal state in non-doped and F-doped LaFePO, respectively. The open triangles, solid squares, and open diamonds correspond to heat capacity measurements of F-doped LaFePO under magnetic fields of 1, 2, and 3 T, respectively. The other symbols are the same as those in Fig. 1. The inset displays the same plot near the Schottky anomaly.

**Fig. 3.** Electronic heat capacity per unit temperature ($C_{ele} T^{-1}$) of F-doped LaFePO below 8 K. The dotted and dashed curves show an entropy-conserving construction, which is used to determine $\Delta C_p/T_c$. The symbols are the same as those in Fig. 1.

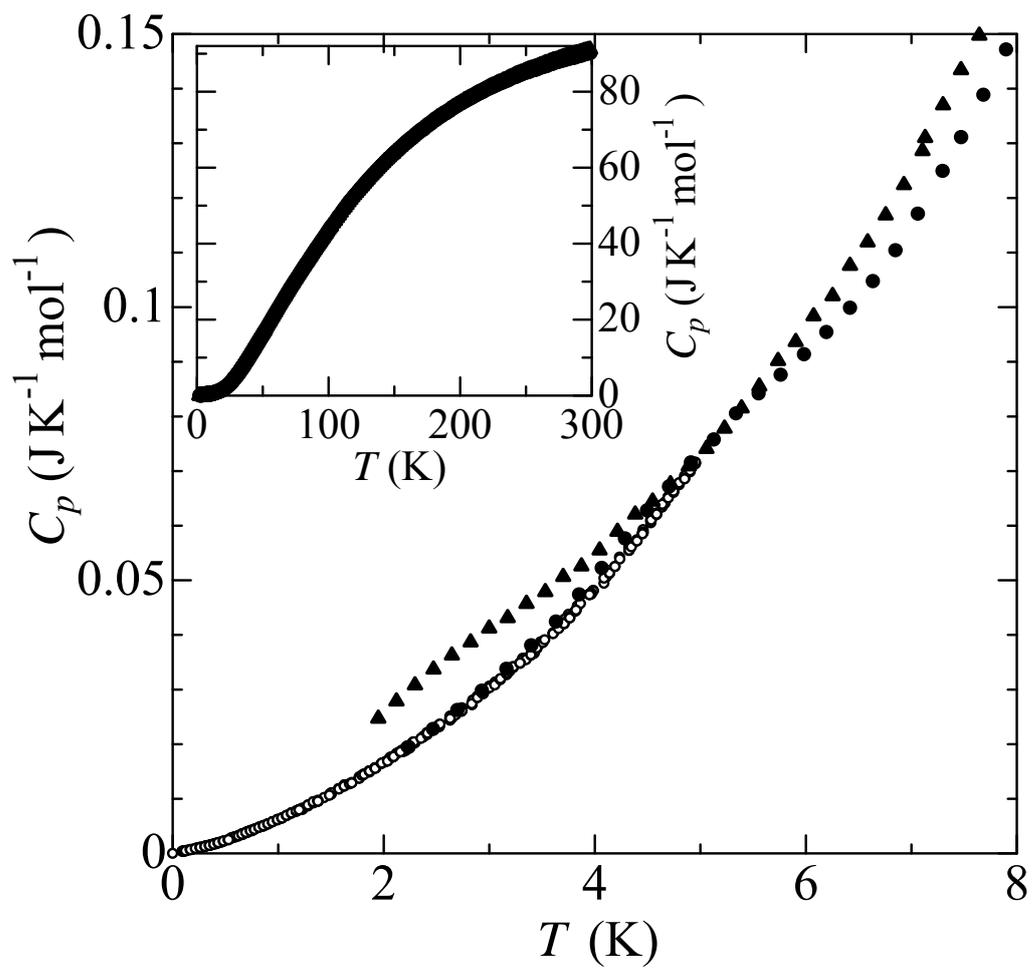

Fig. 1.

Y. KOHAMA, et al.

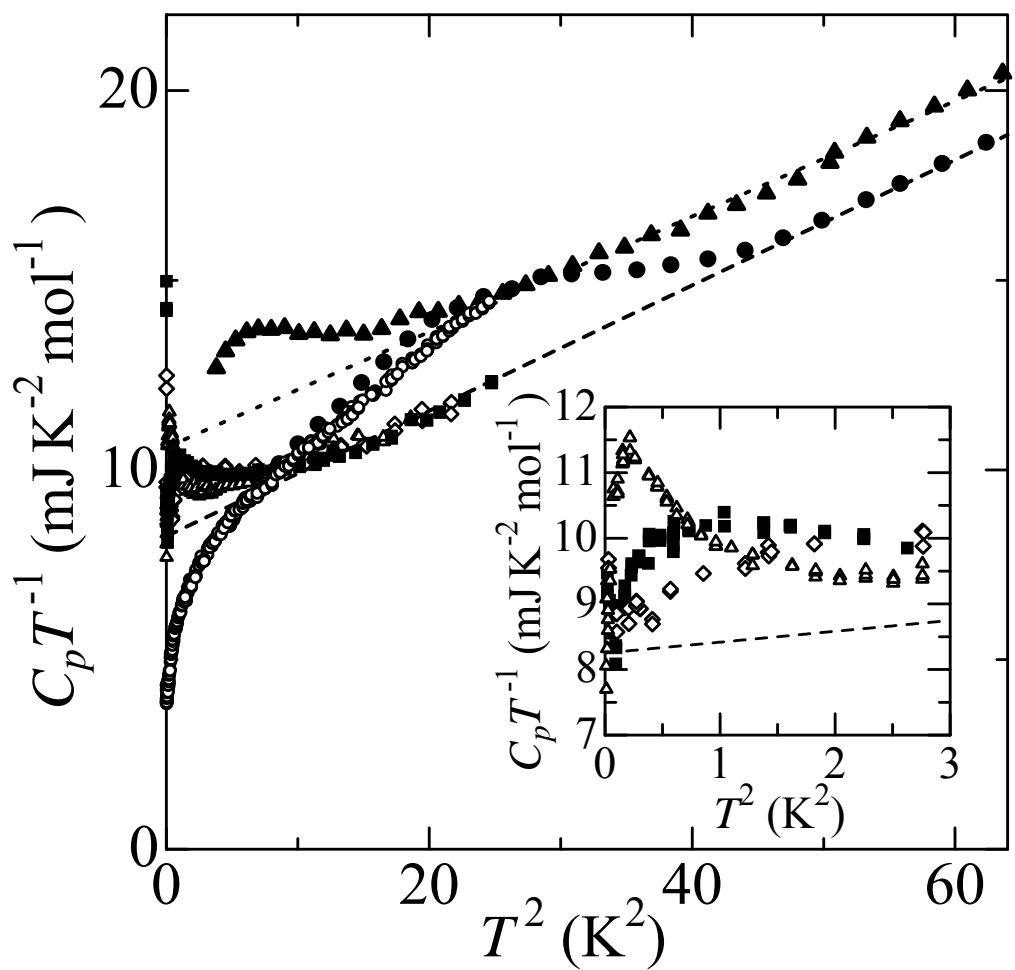

Fig. 2.

Y. KOHAMA, et al.

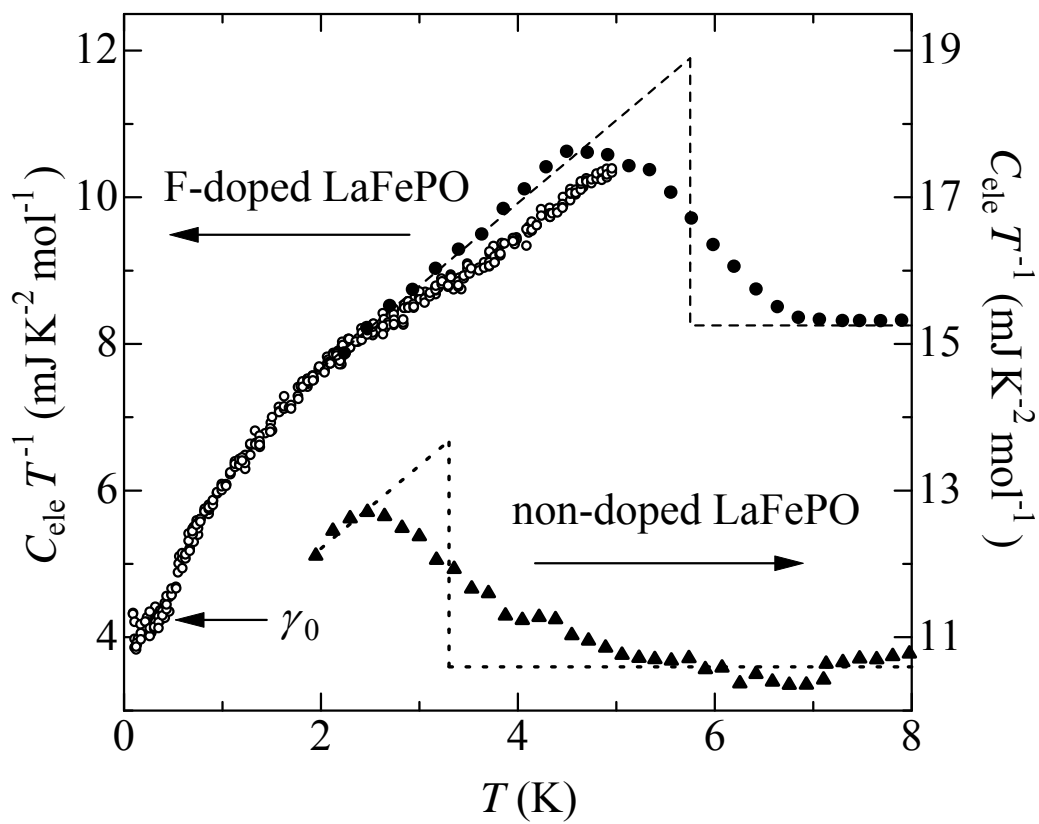

Fig. 3.

Y. KOHAMA, et al.